\documentstyle[12pt]{article}
\textheight
230mm

\def\be{\begin{equation}}
\def\ee{\end{equation}}
\def\bea{\begin{eqnarray}}
\def\eea{\end{eqnarray}}

\begin{document}

\begin{center}
{\Large{\bf Superstring in the Presence of the Worldsheet Gauge Field}}\\
\vskip .5cm 
{\large Davoud Kamani} 
\vskip .1cm 
{\it Faculty of Physics, Amirkabir University of Technology 
(Tehran Polytechnic)\\ 
P.O.Box: 15875-4413, Tehran, Iran}\\
{\sl e-mail: kamani@aut.ac.ir}
\\
\end{center}

\begin{abstract}

We study the superstring theory with an Abelian
worldsheet gauge field. The components of the gauge field appear
as a space and a time coordinates. We call them as ``fictitious
coordinates''. The worldsheet supersymmetry and the
Poincar\'e symmetry of this model will be analyzed. The
$T$-duality and quantization of the two fictitious coordinates also
will be obtained.

\end{abstract}
\vskip .5cm



\newpage
\section{Introduction}
The gauged actions in the string theory have been studied from the
various point of view \cite{1}. We introduce
an Abelian worldsheet gauge field
in the superstring action. Our motivation is as follows.
The S-duality of the type IIB theory implies that fundamental
string of IIB is S-dual to the D-string. Thus, according
to the observation of \cite{2}, the worldsheet fields
comprise not only the target space coordinates, but also
an $SL(2;R)$ doublet of Abelian gauge fields.
In addition, the worldsheet with gauge field in the Matrix string
theory also has been studied \cite{3}.

We are interested in the case that
the $U(1)$ gauge field, which lives on the string worldsheet, to be
independent degree of freedom. Therefore, the square form of the
corresponding field strength appears in the string action. 
The gauged action of the superstring enables us to build two
worldsheet fields from the components of the gauge field. They appear
as the space and time coordinates. Since they are not actual
coordinates, we call them as ``{\it fictitious coordinates}''.
The fictitious space coordinate is hidden. 
In other words, from the quantization
point of view, we observe that it has to be compact. We study the
various symmetries, $T$-duality and quantization of this extended
action of the superstring.

This paper is organized as follows. In section 2, the superstring
action in the presence of the worldsheet 
gauge field $A_a(\sigma, \tau)$ will be
obtained. In section 3, the components of the field $A^a$ as
the fictitious coordinates, and also their super-partners will be studied. In
section 4, some symmetries of the model
will be discussed. In section 5, the solutions of the fictitious
coordinates and some electric fields will be obtained. In section
6, the effects of $T$-duality on the fictitious coordinates and also
their quantizations will be analyzed.

\section{The gauged action of the superstring }

The action of the superstring with the worldsheet supersymmetry is
\bea 
S_0 = -\frac{1}{4\pi \alpha'}\int d^2 \sigma (\partial^a
X^\mu
\partial_a X_\mu-i{\bar \psi}^\mu \rho^a \partial_a \psi_\mu),
\eea
where we have the following notations
\bea
\rho^0=\left(\begin{array}{cc}
0 & -i \\
i & 0
\end{array}\right)\;\;\;,
\;\;\;\;\;
\rho^1=\left(\begin{array}{cc}
0 & i \\
i & 0
\end{array}\right),
\eea 
for the matrices $\{ \rho^a\}$ in the Majorana basis, and
$\eta_{ab}={\rm diag}(-1,1)$ and $\eta_{\mu\nu}={\rm
diag}(-1,1,...,1)$ for the worldsheet and spacetime metrics. The
supersymmetry transformations for this action are 
\bea 
&~& \delta X^\mu = {\bar \eta} \psi^\mu,
\nonumber\\
&~& \delta \psi^\mu = -i \rho^a \partial_a X^\mu \eta, 
\eea 
where $\eta$ is a constant infinitesimal Majorana spinor. The
supercurrent, associated to these transformations, is 
\bea
j_a=\frac{1}{2}\rho^b\rho_a\psi^\mu\partial_b X_\mu, 
\eea 
which is a conserved current, $i.e.$ $\partial^a j_a=0$.

To obtain a supersymmetric action with the gauge field, we use the
superfields in the worldsheet superspace. In other words, the
bosonic action 
\bea 
S_1 = -\int d^2\sigma \bigg{(}\frac{1}{4\pi
\alpha'}\partial_a X^\mu
\partial^a X_\mu +\frac{1}{4g^2} F_{ab}F^{ab}\bigg{)},
\eea 
should be supersymmetrized. The factor $g$ is gauge coupling constant
and the field strength is $F_{ab}=\partial_a A_b-\partial_b A_a$.
This action has the gauge
symmetry. Therefore, we have the condition 
\bea
\partial_a A^a=0.
\eea 
Note that the worldsheet field $A_a$ is not pull-back of a
spacetime gauge field $A_\mu$. 

The bosonic fields $X^\mu$ and $A^a$ should be replaced by the
superfields 
\bea 
Y^\mu(\sigma, \tau;\theta^1,\theta^2)=X^\mu(\sigma, \tau)+{\bar
\theta}\psi^\mu(\sigma, \tau)+\frac{1}{2}{\bar \theta}\theta
B^\mu(\sigma, \tau), 
\eea 
\bea 
{\cal{A}}^a(\sigma,
\tau;\theta^1,\theta^2)=A^a(\sigma, \tau)+{\bar \theta}\rho^a
\chi(\sigma, \tau)+\frac{1}{2}{\bar \theta}\theta W^a(\sigma,\tau), 
\eea 
where $B^\mu$ and $W^a$ are auxiliary fields and the Majorana spinor
$\chi$ is the super-partner of $A_a$. The
Grassmannian coordinates $\theta^1$ and $\theta^2$ form a Majorana
spinor $\theta=\left(\begin{array}{c}
\theta^1 \\
\theta^2
\end{array}\right)$.
Thus, the term such as ${\bar \theta}\rho^a \theta u(\sigma ,
\tau)$ is zero and hence does not appear in the superfield (8).

The derivative $\partial_a$ also should be changed. We
introduce the following superspace covariant derivative 
\bea 
&~& {\cal{D}}^a=k \varepsilon^{ab} \rho_b D,
\nonumber\\
&~& D=\frac{\partial}{\partial{\bar \theta}}-i\rho^a \theta
\partial_a,
\eea 
where $\varepsilon^{01}=-\varepsilon^{10}=1$ and $k$ is a constant which 
will be specified. For a
superfield $Z$ its derivative $DZ$ also is  superfield. On the
other hand, the transformation $\delta Z={\bar \epsilon}Q(Z)$
implies $\delta(DZ)={\bar \epsilon}Q(DZ)$. The operator 
\bea
Q=\frac{\partial}{\partial{\bar \theta}}+i\rho^a \theta
\partial_a,
\eea 
is generator of the supersymmetry on the superspace.
Similarly, ${\cal{D}}^a Z$ also is a superfield. That is, it
transforms in the same way 
\bea 
\delta({\cal{D}}^a Z)={\bar
\epsilon}Q({\cal{D}}^a Z). 
\eea 
Therefore, the derivative
${\cal{D}}^a$ enables us to write the supersymmetric form of the
action (5). For finding the constant $k$ we demand the property
\bea 
{\bar {\cal{D}}}^a Y_\mu{\cal{D}}_a Y^\mu={\bar D} Y_\mu D
Y^\mu. 
\eea 
This gives $k\in \{\pm\frac{1}{{\sqrt 2}} ,
\pm\frac{i}{{\sqrt 2}}\}$. Note that the definition ${\cal{D}}^a
= k \rho^a D$ does not satisfy the equation (12).

Adding all these together we obtain the following supersymmetric
action 
\bea 
S = \int d^2 \sigma d^2\theta \bigg{(}\frac{i}{8\pi
\alpha'}{\bar {\cal {D}}}^a Y_\mu{\cal {D}}_a Y^\mu +
\frac{i}{4g^2} {\bar {\cal{F}}}_{ab}{\cal{F}}^{ab}\bigg{)}. 
\eea
The superfield strength ${\cal{F}}_{ab}$ has the definition 
\bea
{\cal{F}}_{ab}={\cal{D}}_a{\cal{A}}_b-{\cal{D}}_b{\cal{A}}_a. 
\eea
After integration over the Grassmannian coordinates $\theta^1$ and
$\theta^2$, this action takes the form 
\bea 
S =\int d^2\sigma
\bigg{(}-\frac{1}{4\pi \alpha'}(\partial_a X^\mu
\partial^a X_\mu -i{\bar \psi}^\mu \rho^a \partial_a
\psi_\mu-B^\mu B_\mu) -\frac{1}{2g^2} (\partial_a A^b\partial^a
A_b - W_aW^a)\bigg{)}. 
\eea 
As we see, the gagino field $\chi$
from the two-dimensional action disappeared. 
In addition, according to the
condition (6), the term $\partial_a A^b\partial_b A^a$ is double
total derivative $\partial_a\partial_b(A^aA^b)$ and hence it also has
been ignored.
\section{A new form for the gauged action}

In the action (15) the kinetic terms of the fields $X^\mu$ and
$A^b$ have the same feature. In other words, $A^0$ and $A^1$ have
the roles of the time and space coordinates.
Let $\{X^a\}$ denote the coordinates of this fictitious 1+1 dimensional
spacetime. Thus, we have the field redefinition 
\bea 
X^a =\frac{\sqrt{2\pi\alpha'}}{g}A^a . 
\eea 
In addition, the auxiliary
field $B^a$ also is scaled in the same way 
\bea
B^a=\frac{\sqrt{2\pi\alpha'}}{g}W^a. 
\eea 
According to these
definitions, the action (15) can be written as 
\bea 
S=-\frac{1}{4\pi \alpha'}\int d^2\sigma (\partial_a X^M
\partial^a X_M -i{\bar \psi}^\mu \rho^a \partial_a \psi_\mu-B^M B_M),
\eea 
where $M\in \{\mu , a\}$, and we shall use the convention 
$a\in \{0,1\}$ and $\mu \in\{0',1',...,9'\}$.
Since both $X^a$ and $\sigma^a$ carry the worldsheet index,
the partial derivative $\partial_a$ always shows derivative with
respect to $\sigma^a$. The bosonic part of this action 
apparently describe
a 12-dimensional spacetime with the signature 10+2 and the
coordinates 
\bea 
\{X^M\}=\{X^\mu\}\bigcup\{X^a\}.
\eea 
Because the gauge condition is survived, 
this two additional coordinates are constrained on the worldsheet.
Thus, they do not describe additional 1+1 dimensions.
However, in the superstring theory the dimension of the
spacetime always is 9+1. Therefore, we call the extra dimensions as
``fictitious coordinates''.

The fermionic term of the action (18) also can be written with the
12-dimensional indices. For this, the Majorana spinor $\psi^a$ is
defined by
\bea 
\psi^0=\frac{\sqrt{2\pi\alpha'}}{g}\left(\begin{array}{c}
\chi_2 \\
\chi_1
\end{array}\right)
\;\;\;\;,\;\;\;\; \psi^1=
\frac{\sqrt{2\pi\alpha'}}{g}\left(\begin{array}{c}
\chi_2 \\
-\chi_1
\end{array}\right).
\eea
The spinors $\psi^0$ and $\psi^1$ satisfy the identities 
\bea
{\bar \psi}^b \rho^a\partial_a \psi_b=0, 
\eea 
\bea {\bar \psi}^a
\psi_a=\frac{4\pi\alpha'}{g^2}{\bar \chi}\chi, 
\eea 
where
$\chi=\left(\begin{array}{c}
\chi_1 \\
\chi_2
\end{array}\right)$.
Introducing the identity (21) in the action (18) leads to the covariant form
of this action 
\bea 
I =-\frac{1}{4\pi\alpha'}\int d^2\sigma (\partial_a X^M
\partial^a X_M -i{\bar \psi}^M \rho^a \partial_a \psi_M-B^M B_M).
\eea 
The metric of the extended manifold is 
\bea 
\eta_{MN}={\rm diag}(\eta_{\mu\nu} , \eta_{ab}), 
\eea
where $\eta_{\mu\nu}$ belongs to the 9+1 actual spacetime and 
$\eta_{ab}$ for the fictitious coordinates.

The equations of motion, extracted from the action (23), are 
\bea
&~& \partial_a\partial^a X^M=0,
\nonumber\\
&~& \rho^a \partial_a \psi^M=0,
\nonumber\\
&~& B^M=0.
\eea
In addition, we should also consider the gauge condition
\bea
\partial_a X^a =0.
\eea 
This condition and the equation of motion of $X^a$ can be
written as 
\bea 
X^a = \varepsilon^{ab} \partial_b \phi , 
\eea 
\bea
\partial_a\partial^a \phi= c,
\eea
respectively. The constant $c$ is independent of $\sigma$ and $\tau$.

The fields in the right-hand-sides of the equations (16), (17)
and (20) correspond to the gauge theory, while the fields in the
left-hand-sides belong to the string theory. The factor
$\frac{\sqrt{2\pi\alpha'}}{g}$ in these relations, for the weak
coupling regime in the gauge part, $i.e.$ $g \rightarrow 0$, is
equivalent to the zero string tension $\alpha' \rightarrow
\infty$. This is an expected result from the duality map between
gauge theories and strings \cite{4}.
\section{Symmetries of the model}
\subsection{Worldsheet supersymmetry}

Using the superfield (8) we obtain the supersymmetry
transformations of $A^a$ and $\chi$ as in the following 
\bea 
&~&\delta A^a = {\bar \epsilon}\rho^a\chi,
\nonumber\\
&~& \delta \chi=-\frac{i}{4}\rho^{ab} F_{ab} \epsilon -\frac{1}{2}
\rho_a W^a \epsilon,
\nonumber\\
&~& \delta W^a = -i {\bar \epsilon}\rho^b \rho^a \partial_b \chi,
\eea 
where $\rho^{ab}=\frac{1}{2}[\rho^a , \rho^b]$. The
supersymmetry parameter $\epsilon$ is an anti-commuting
infinitesimal constant spinor. For obtaining these
transformations the gauge condition (6) has been used. In terms
of the fields $\{X^a , \psi^a , B^a\}$ these transformations take
the form 
\bea 
&~& \delta X^a = i\varepsilon^{ab} {\bar
\epsilon}\psi_b ,
\nonumber\\
&~& \delta \psi^a =-\frac{1}{2}(\rho^a\varepsilon^{bc}\partial_b X_c
-i\varepsilon^{ab}\rho_b\rho_c B^c)\epsilon ,
\nonumber\\
&~& \delta B^a = \varepsilon^{ab} {\bar \epsilon}\rho^c \partial_c \psi_b .
\eea

The transformations (30) form a closed algebra. To see this, use
the equations of motion to remove $B^a$ from these
transformations. Now consider two successive transformations with
the supersymmetry parameters $\epsilon_1$ and $\epsilon_2$; then
\bea 
[\delta_{\epsilon_1} , \delta_{\epsilon_2}] X^a = \Omega^a
\varepsilon^{a'b'}
\partial_{a'}X_{b'} ,
\eea 
for the worldsheet bosons $\{X^a\}$, where $\Omega^a =
i\varepsilon^{ab}{\bar \epsilon_1}\rho_b \epsilon_2$. For the
worldsheet fermions $\{\psi^a\}$ we have 
\bea
[\delta_{\epsilon_1} , \delta_{\epsilon_2}] \psi^a = i{\bar
\epsilon_1}\rho^b \epsilon_2
\partial_b \psi^a. 
\eea

The supercurrent associated to the supersymmetry transformations
(30), accompanied by $\delta X^\mu=\delta \psi^\mu =0$, is 
\bea 
k_a = \frac{i}{4} \rho^{a'}\rho_a \psi_{a'}
\varepsilon^{bc}\partial_b X_c .
\eea 
According to the identity $\rho^a\rho^{a'}\rho_a=0$ there is
$\rho^a k_a=0$. That is, some light-cone components of $k_a$
vanish. The field equations (25) 
for $M\in \{0,1\}$, and the gauge condition (26)
imply that $k_a$ is a conserved current, $i.e.$, 
\bea
\partial^a k_a=0.
\eea 
The transformations (3) and (30) for $\eta = \epsilon$ give
the conserved current $J_a =j_a+k_a$.

It is possible to define a two-index current $K_{ab}$ as in the
following 
\bea 
K_{ab} = \frac{i}{4}\rho_a \psi_b
\varepsilon^{a'b'}\partial_{a'} X_{b'} .
\eea 
In fact, we have $k_a=\rho^aK_{ab}$. 
In terms of the fields $A_a$ and $\chi$ it becomes 
\bea
K_{ab}=\frac{\pi \alpha'}{4g^2}\rho_a\rho^c \chi \varepsilon_{bc}
\varepsilon^{a'b'}F_{a'b'}. 
\eea
This current also satisfies the conservation law 
\bea
\partial^a K_{ab}=0.
\eea 

The action (23) also is invariant under the following
supersymmetry transformations 
\bea 
&~& \delta X^M = {\bar \kappa}\psi^M ,
\nonumber\\
&~& \delta \psi^M =-i\rho^a \partial_a X^M \kappa +B^M \kappa ,
\nonumber\\
&~& \delta B^M = -i{\bar \kappa}\rho^a \partial_a \psi^M, 
\eea
where $\kappa$ is the supersymmetry parameter. The associated
conserved current is 
\bea 
{\cal{J}}_a =\frac{1}{2} \rho^b \rho_a\psi^M\partial_b X_M . 
\eea
\subsection{Consistency of the gauge condition with the supersymmetry}

The gauge condition (26) can also be written in the form
\bea
\partial_+ X^0 +\partial_+ X^1 +\partial_- X^0 -\partial_- X^1=0,
\eea
where $\partial_{\pm}=\frac{1}{2}(\partial_\tau \pm \partial_\sigma)$.
According to the worldsheet supersymmetry, we should also introduce
the fermionic analog of (40), $i.e.$,
\bea
\psi^0_+ +\psi^1_+ +\psi^0_- -\psi^1_- =0,
\eea
where $\psi^a_{\pm}$ are defined by 
$\psi^a = \left(\begin{array}{c}
\psi^a_-\\
\psi^a_+
\end{array}\right)$.
According to (20) $\psi^a_-$ ($\psi^a_+$) has expression in terms of $\chi_2$
($\chi_1$). Therefore, the equation (41) is an
identity. That is, the gauge condition (26) is consistent with 
supersymmetry.
\subsection{ The Poincar\'e symmetry}

The action (23), with $B^M=0$, under the Poincar\'e transformations
\bea
&~& \delta X^M = a^M_{\;\;\;N}X^N + b^M ,
\nonumber\\
&~& \delta \psi^M =a^M_{\;\;\;N} \psi^N ,
\eea
is symmetric. The matrix $a_{MN}$ is constant and antisymmetric, and
$b^M$ is a constant vector. The associated currents to these
transformations are
\bea
&~& P^M_a = \frac{1}{2\pi \alpha'} \partial_a X^M ,
\nonumber\\
&~& J^{MN}_a = \frac{1}{2\pi \alpha'} (X^M \partial_a X^N -X^N
\partial_a X^M +i{\bar \psi}^M\rho_a \psi^N). 
\eea 
These are conserved currents, $i.e.$ $\partial^a P^M_a=\partial^a J^{MN}_a=0$.

According to the gauge condition (26)
the component $J^{bc}_a$ takes the form 
\bea
J^{bc}_a=\frac{1}{2\pi \alpha'}
(-\varepsilon^{bc}\varepsilon_{aa'} X^{b'} \partial_{b'} X^{a'}
+i {\bar \psi}^b\rho_a \psi^c).
\eea 
\section{The fictitious coordinates and electric fields}
\subsection{Solutions of the fictitious coordinates}

Now we solve the equation of motion 
\bea 
(\partial^2_\tau -\partial^2_\sigma)X^a=0,
\eea 
for the cases that the fictitious coordinates $\{X^a(\sigma, \tau)|a=0,1\}$ 
appear as the coordinates of
open and closed strings. The variation of the action
(23) leads to the following boundary condition for open string solution
\bea 
(\partial_\sigma X^a)_{\sigma_0}=0, 
\eea 
where $\sigma_0 =0, \pi$. Therefore, the solutions of the equation (45) are 
\bea
X^a(\sigma , \tau)= x^a + l^2 p^a \tau + il \sum_{n \neq 0}
\frac{1}{n} \alpha^a_n e^{-in\tau} \cos(n \sigma) , 
\eea 
for the open string, and 
\bea 
X^a(\sigma , \tau)= x^a + l^2 p^a \tau + 2L^a
\sigma +\frac{i}{2}l \sum_{n \neq 0} \frac{1}{n}
\bigg{(}\alpha^a_n e^{-2in(\tau -\sigma)}+ {\tilde \alpha}^a_n
e^{-2in(\tau +\sigma)}\bigg{)}, 
\eea 
for the closed string. In these
solutions there are $l=\sqrt{2\alpha'}$ and $L^a=n^a R_a$, where
$n^a$ is winding number of the closed string and $R_a$ is radius of
compactification of the fictitious coordinate $X^a$.

These solutions should also satisfy the condition (26). This
condition gives 
\bea 
p^0=\alpha^0_n=\alpha^1_n=0 , 
\eea 
for the open string, and
\bea
&~& l^2 p^0 +2 L^1 =0 ,
\nonumber\\
&~& \alpha^1_n=\alpha^0_n \equiv \alpha_n ,
\nonumber\\
&~& {\tilde \alpha}^1_n=-{\tilde \alpha}^0_n \equiv -{\tilde
\alpha}_n , 
\eea 
for the closed string. Introducing (49) and (50) into
the solutions (47) and (48) we obtain 
\bea 
&~& X^0(\sigma ,\tau)=x^0,
\nonumber\\
&~& X^1(\sigma , \tau)=x^1 + l^2 p^1 \tau ,
\eea
as fictitious coordinates of the open string, and
\bea
&~& X^0(\sigma , \tau)= x^0 + 2(L^0 \sigma - L^1 \tau)
+\frac{i}{2}l \sum_{n \neq 0} \frac{1}{n}
\bigg{(}\alpha_n e^{-2in(\tau -\sigma)}+
{\tilde \alpha}_n e^{-2in(\tau +\sigma)}\bigg{)},
\nonumber\\
&~& X^1(\sigma , \tau)= x^1 + l^2(p^1 \tau- p^0 \sigma)
+\frac{i}{2}l \sum_{n \neq 0} \frac{1}{n}
\bigg{(}\alpha_n e^{-2in(\tau -\sigma)}-
{\tilde \alpha}_n e^{-2in(\tau +\sigma)}\bigg{)},
\eea
for the closed string.
\subsection{Various electric fields}

The worldsheet field strength
$F_{ab}=\frac{g}{\sqrt{2\pi\alpha'}}(\partial_aX_b-\partial_bX_a)$
can be written as $F_{ab}=-\varepsilon_{ab}E$, where the worldsheet
electric field $E$ is 
\bea 
E=\frac{gl^2p^1}{\sqrt{2\pi\alpha'}},
\eea 
from the open string solution (51), and 
\bea
E=\frac{g(l^2p^1+2L^0)}{\sqrt{2\pi\alpha'}}, 
\eea 
due to the closed string solution (52). These imply that the zero modes of
the fictitious coordinates specify the field strength
on the worldsheet.

Now we prove that the compactness of the coordinate $X^1$ imposes
an electric field along it. There is the following relation
between the momentum and winding numbers of a closed string
\cite{5}, 
\bea 
p^a =-\frac{1}{\alpha'}f^a_{\;\;\;a_c}L^{a_c},
\eea 
where the index ${a_c}$ refers to the compact
directions and $f_{ab}$ is a background field
strength. Since the $X^1$-direction is compact this relation
gives $p^0 = -\frac{1}{\alpha'}f^0_{\;\;\;1}L^1$. The
other form of this equation is 
\bea 
p^0 =\frac{1}{\alpha'}{\cal{E}}L^1, 
\eea 
where the electric field along $X^1$-direction is defined 
by ${\cal{E}} \equiv f_{01}$. The
equation (56) with the first equation of (50) lead to 
\bea
{\cal{E}}=-1. 
\eea 
Therefore, there is a unit electric field in
the $X^1$-direction. Note that ${\cal{E}} \neq F_{01}$.
\section{T-duality and quantization of the fictitious coordinates}
\subsection{T-duality}

The zero modes $x^a$, $p^a$ and $L^a$, in terms of the left and
right components are 
\bea 
&~& x^a=x^a_L+x^a_R,
\nonumber\\
&~& p^a=p^a_L+p^a_R,
\nonumber\\
&~& L^a=\alpha'(p^a_L-p^a_R). 
\eea 
Under the $T$-duality $x^a_L$,
$p^a_L$ and ${\tilde \alpha}_n$ do not change, while $x^a_R$,
$p^a_R$ and $\alpha_n$ change their signs. Thus, the $T$-duality
on the fictitious coordinates (52) gives 
\bea 
&~& X'^0(\sigma , \tau)= x'^0
+ l^2(p^0 \sigma - p^1 \tau) +\frac{i}{2}l \sum_{n \neq 0}
\frac{1}{n} \bigg{(}-\alpha_n e^{-2in(\tau -\sigma)}+ {\tilde
\alpha}_n e^{-2in(\tau +\sigma)}\bigg{)},
\nonumber\\
&~& X'^1(\sigma , \tau)= x'^1 + 2(L^1 \tau - L^0 \sigma)
-\frac{i}{2}l \sum_{n \neq 0} \frac{1}{n}
\bigg{(}\alpha_n e^{-2in(\tau -\sigma)}+
{\tilde \alpha}_n e^{-2in(\tau +\sigma)}\bigg{)},
\eea
where $x'^a=x^a_L-x^a_R$.

Assume the following relations between the center of mass
coordinates $x^a$ and $x'^a$, 
\bea 
x'^0 =- x^1\;\;,\;\;x'^1=-x^0 .
\eea 
According to this, compare the dual coordinates (59) with
the coordinates (52). We observe that 
\bea 
X'^0 =-X^1\;\;,\;\;X'^1=-X^0 . 
\eea 
That is, up to the sign, the
fictitious coordinates are $T$-dual of each other. As expected,
under twice dualization these coordinates do not change. The
compact form of (61) is 
\bea 
X'^a = -\varepsilon^{ab} X_b . 
\eea
Combine this with the equation (27). In terms of the field $\phi$
this dual coordinate is 
\bea 
X'^a = -\eta^{ab}\partial_b \phi . 
\eea 
The equations (28) and (63) give $\partial_a
X'^a=-c$. Therefore, the dual coordinates also are constrained
by this gauge condition.
However, the equations (27) and (63) imply that the
scalar field $\phi$ is origin of the four coordinates $\{X^0,
X^1;X'^0, X'^1\}$.

Define the divergence of the gauge field as $\Phi = -\partial_a A^a$.
From the equations (61) we observe that under the $T$-duality
$\Phi$ and $F_{01}=\frac{g}{\sqrt{2\pi\alpha'}}(\partial_\tau
X^1+\partial_\sigma X^0)$ get exchange 
\bea 
\Phi\longleftrightarrow F_{01}. 
\eea 
In other words, $T$-dual of
the field strength is zero. The $T$-dual version of $F_{ab}$ also
can be obtained from (54) 
\bea
F'_{ab}=-\varepsilon_{ab}\frac{g(2L^1+l^2
p^0)}{\sqrt{2\pi\alpha'}}. 
\eea 
The first equation of (50) gives
$F'_{ab}=0$. This is consistent with the $T$-duality relation (64).
\subsection{Quantization of the extra worldsheet fields}

Since the condition (41) is an identity
the worldsheet fermions $\{\psi^a\}$ are not 
constrained. Thus, in their quantizations there are not new
phenomena. For quantizing the bosonic fields $\{X^a\}$ we need to
fix the gauge. The gauge fixed Lagrangian is 
\bea
{\cal{L}}=-\frac{1}{4\pi \alpha'}(\partial_a X^b\partial^a X_b +
\lambda (\partial_a X^a)^2), 
\eea 
where $\lambda$ is constant. In
addition to the equation of motion, extracted from this
Lagrangian, we should also consider the equation (26),
$e.g.$ see \cite{6}. The result
is the same as the case $\lambda = 0$.

Since quantization does not admit the open string solution (51),
we consider the closed string solution (52). The canonical
momentum conjugate to $X^a$ is 
\bea 
\Pi^a(\sigma ,\tau)=\frac{1}{2\pi \alpha'}\partial_\tau X^a . 
\eea 
The canonical commutation relation at equal $\tau$ is 
\bea 
[X^a(\sigma, \tau) , \Pi^b(\sigma' , \tau)] = i\delta^{ab} \delta(\sigma
-\sigma') . 
\eea 
In fact, consistency with the constraint (26)
requires $\delta^{ab}$ instead of $\eta^{ab}$. This quantization
leads to the following commutation relations 
\bea 
&~& [x^0 ,p^0]=[x^1 , p^1]=i,
\nonumber\\
&~& [\alpha_m , \alpha_n]=[{\tilde \alpha}_m , {\tilde \alpha}_n]=
m\delta_{m+n , 0},
\nonumber\\
&~& [\alpha_m , {\tilde \alpha}_n]=0.
\eea

According to the first equation of (50), the commutator $[x^0 ,
p^0]=i$ implies that $L^1 \neq 0$. In other words, the worldsheet
field $X^1$ always is compact. 
The worldsheet field $X^0$ can be compact or non-compact.

The following nontrivial commutation relations are consequences
of the relations (69), 
\bea 
&~& [X^0(\sigma , \tau) , X^1(\sigma' ,
\tau)] = 2i \alpha' \sum^\infty_{n=1}\frac{\sin [2n(\sigma
-\sigma')]}{n} , 
\nonumber\\
&~& [\Pi^0(\sigma , \tau) , \Pi^1(\sigma'
, \tau)] = -\frac{i}{2\pi \alpha'}\delta'(\sigma -\sigma') , 
\eea
where $\delta'(\sigma -\sigma')=\partial_\sigma\delta(\sigma
-\sigma')$. These commutators do not show the usual
noncommutativity. For the usual noncommutativity we should
consider $\sigma = \sigma'$, which gives zero for the
right-hand-sides of (70). However, the fictitious
coordinates in different points of closed string have a kind of
noncommutativity.
\section{Conclusions}

By appropriate definitions for the superfields and covariant
derivative on the worldsheet superspace, we studied the
superstring theory in the presence of a worldsheet gauge field.
This model contains
some connections between the string and gauge fields. For the
fictitious 1+1 directions, the worldsheet supersymmetry and the
associated currents were obtained. In addition, the Poincar\'e
symmetry was analyzed.

The fictitious open string coordinates only have zero modes. For
the closed string, the winding number around the $X^1$-direction
is proportional to the momentum along the $X^0$-direction. This
gives a unit electric field along the
$X^1$-direction. However, the worldsheet electric fields have
expressions in terms of the momentum and winding number of the
string.

We observed that the fictitious closed 
string coordinates $X^0$ and $X^1$ are
$T$-dual of each other. In addition, the field strength and the divergence
of the gauge field, under the $T$-duality, transform to each
other. Therefore, $T$-dual of the field strength vanishes. A
worldsheet scalar $\phi (\sigma, \tau)$ defines the 
fictitious coordinates $X^0$ and $X^1$
and their $T$-dual coordinates.

We saw that the quantization of the field $X^0$ imposes the
compactification of the field $X^1$. Finally, in the
space $\{X^0, X^1\}$ we obtained a noncommutativity between
the coordinates of two different points of closed string.

\end{document}